\documentclass{elsart}

\usepackage{natbib}
\bibpunct{(}{)}{;}{a}{}{,}

 \usepackage{graphics}
 \usepackage{graphicx}
 \usepackage{epsfig}

\usepackage{amssymb}

\def\ie {$i.e.,\;$}

\def\ie {$i.e.,\;$}

\def\La1215 {Ly$\alpha\lambda1215$}

\begin{document}
\begin{frontmatter}

\title{Pc-scale study of Radio galaxies \& BL~Lacs}
\author[label1]{P. Kharb},
\ead{rhea@iiap.ernet.in}
\author[label1]{P. Shastri},
\author[label2,label3]{D. C. Gabuzda}
\address[label1]{Indian Institute of Astrophysics, Koramangala P.O.,
Bangalore - 560034}
\address[label2]{Physics Department, University College Cork, Cork, Ireland}
\address[label3]{Astro Space Center, Lebedev Physical Institute, Moscow, Russia}

\begin{abstract}
We study two aspects of the differences between Active Galactic Nuclei (AGN)
of the two Fanaroff-Riley types in order to investigate the causes of the 
F--R divide. We (a) contrast the properties of the optical cores in beamed 
and unbeamed AGN of the two types, incorporating Hubble Space Telescope 
measurements of the unbeamed objects, and (b) contrast the nuclear magnetic 
field geometry of the beamed AGN of the two types using Very Long Baseline 
Polarimetry.

\end{abstract}


\end{frontmatter}

\section{Introduction}

Radio-loud AGN are broadly thought of as having two Fanaroff-Riley types, 
with the low radio-luminosity type~I (FRI) objects exhibiting diffuse radio 
jets and the high-luminosity type II (FRII) objects having 
collimated jets and terminal hotspots.
In their discovery paper, \citet{FanaroffRiley74} had found the luminosity 
divide to be at $2\times 10^{26}$ WHz$^{-1}$ at 178~MHz.
In the framework of the standard Unified Scheme (US), FRI and FRII 
radio galaxies represent the parent populations of BL~Lacs and 
radio-loud quasars, respectively \citep{UrryPadovani95}. 
It follows that a ubiquitous obscuring torus is {\it required} 
in the FRII population but not in the FRI population.
Adopting this framework, we will refer to FRI radio galaxies and 
BL~Lacs as the ``FRI population'' and FRII radio galaxies and quasars as 
the ``FRII population.'' To address the origin of the F--R dichotomy, we 
will look at differences in the properties of (i) the pc-scale optical 
cores and (ii) the magnetic ({\bf B}) field geometry in the pc-scale radio 
jets of the two populations (we assume H$_{o}$=75 km s$^{-1}$ Mpc$^{-1}$, 
q$_{o}$=0.5).

\section{The pc-scale optical cores in radio galaxies}
The Hubble Space Telescope ({\it HST}) 
has revealed unresolved optical cores in the centres of a majority
of FRI and FRII radio galaxies \citep{Chiaberge02}.
These authors suggest that the core emission 
in the FRIs and some FRIIs is optical synchrotron radiation.
Here we examine the evidence that the optical core luminosity, $L_{o}$ 
($K$-corrected), is orientation dependent by searching for correlations 
with the radio core prominence, $R_{c} = S_{core}/S_{ext}$ (at an emitted 
frequency of 5 GHz), which is a statistical indicator of relativistic 
beaming \citep{BlandfordKonigl79} and therefore of orientation.
This comparison is plotted in Figs.~1 and 2 for the FRI and FRII 
populations, respectively. For the beamed objects (\ie BL~Lacs 
and quasars) we have taken the optical core luminosity to be the total 
optical luminosity, assuming the core overwhelms the host galaxy emission.  

\begin{figure}[h]
\centerline{
(a)
\includegraphics[width=6.6cm,height=6.6cm]{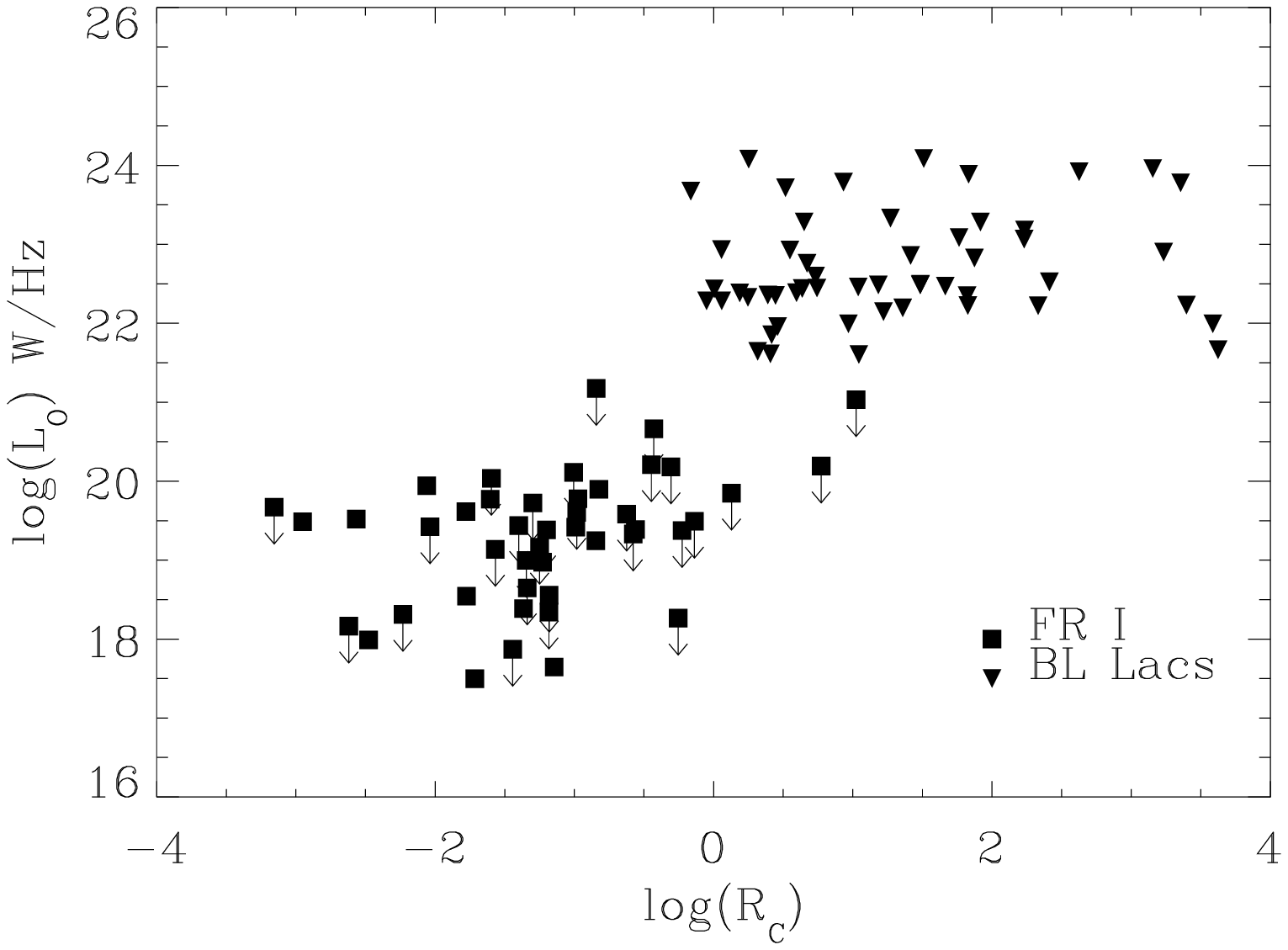}
(b)
\includegraphics[width=6.6cm,height=6.6cm]{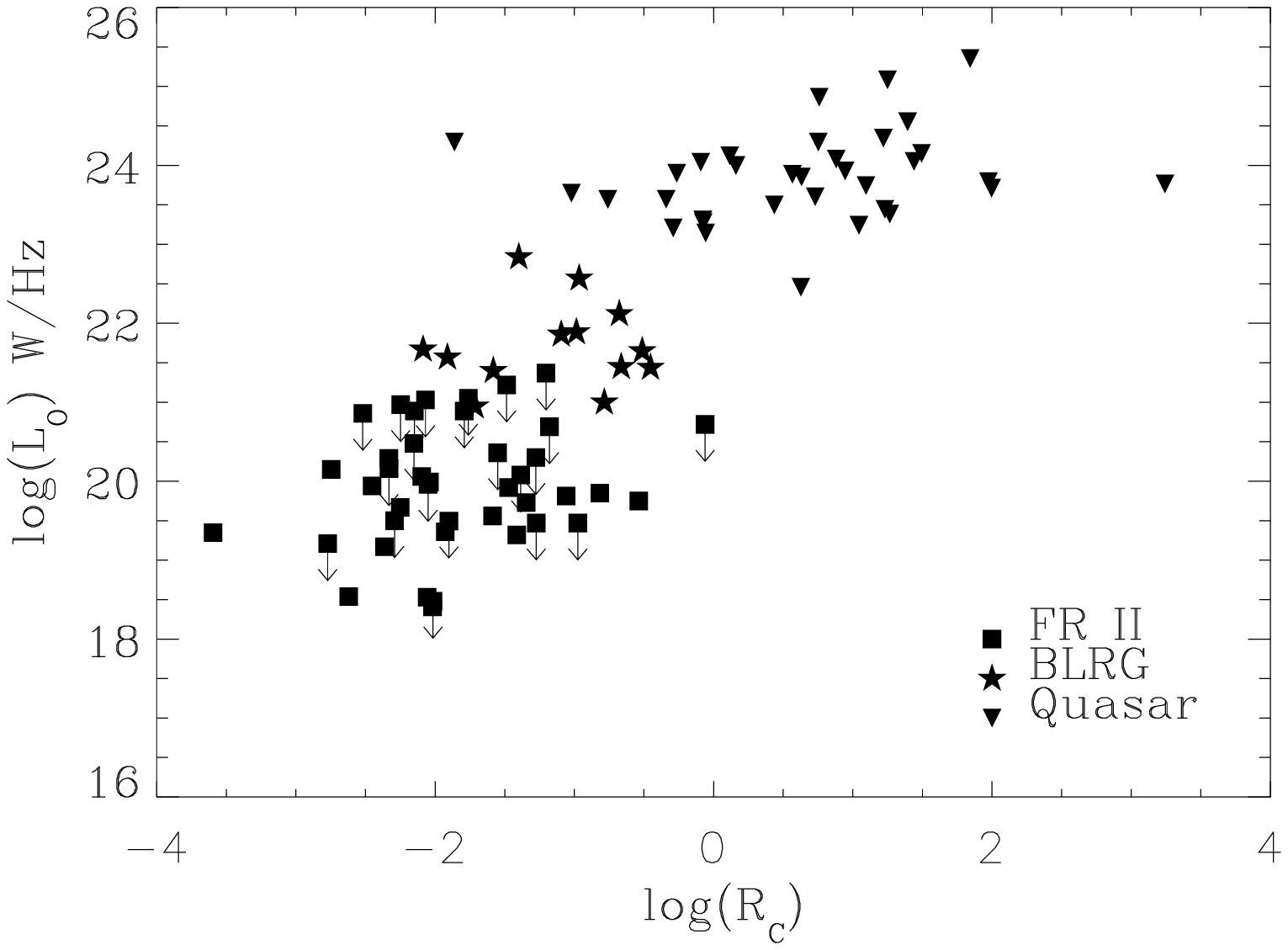}}
\caption{$R_{c}$ vs. optical core luminosity, $L_{o}$, for the (a) FRI 
population and (b) FRII population.  We use $R_{c}$ as a proxy for 
orientation. Arrows denote upper limits.}
\end{figure}

$L_{o}$ is significantly correlated with $R_{c}$ (orientation) for the FRI 
radio galaxies alone ($>$ 0.005 level, Spearman rank correlation test),
as well as for the whole FRI population ($>$ 0.0005 level).  In the FRII 
population, the narrow-lined FRIIs do not show any correlation by themselves,
although a significant ($>$ 0.0005 level) correlation exists for the broad-line 
objects, \ie Broad-Line Radio galaxies (BLRGs) and quasars (as well as for the 
FRII population as a whole).
This is consistent with there being obscuration effects by a torus in
the FRIIs but not in the FRIs.
We note that optical cores have been detected in {\it all} the BLRGs 
observed (where the US predicts no obscuration by the torus), again 
consistent with this idea. 
We attempted to fit a model in which the observed optical 
emission ($L_{o}$) 
is due to contributions from a relativistic optical synchrotron jet 
($L_{jet}$) and a thin accretion disk ($L_{disk}$) using the equations : 

$log L_{o}=log[\delta^{p}*L_{jet}^{int}+L_{disk}*cos(\theta)]$

$log R_{c}=p*log \delta + log R_{c}^{int};$ {\footnotesize{\it int}} =intrinsic  

Here $\delta$ and $\theta$ are the Doppler factor and inclination, 
and $p$ =(2+$\alpha$) or (3+$\alpha$), depending on whether the jet is 
continuous or blobby ($\alpha$ =jet spectral index). 

In the best-fit models, the instrinsic ({\it i.e.,} unbeamed) jets were
only an order of magnitude more luminous in the FRII population, 
while the FRII accretion disks were more luminous than the FRI disks by 
three orders of magnitude.  However, there are some caveats in this 
analysis: (i) the BL~Lacs show a large scatter, which could be due to 
variability, (ii) there is a paucity of FRI objects at intermediate values 
of $R_{c}$, (iii) the $L_{o}$ values for the BL~Lacs and quasars are total 
values, and include the host galaxy contribution, and (iv) the beamed and 
unbeamed objects were not matched in redshift and extended radio luminosity. 

\section{The pc-scale {\bf B} field geometry in BL~Lac objects}
VLBI polarimetry has shown that a majority of radio-loud 
quasars have jet {\bf B} fields aligned with the local  
jet direction, while most BL~Lacs have transverse jet {\bf B} fields 
\citep{Cawthorne93}.  Most of these BL~Lacs were radio-selected 
BL~Lacs (RBLs); subsequently, many more BL~Lacs were discovered in X-ray 
surveys (X-ray selected BL Lacs, or XBLs). RBLs show greater radio core 
prominence, variability and optical polarization than XBLs 
\citep{LaurentMuehleisen93}, 
suggesting that RBLs are more beamed than XBLs.
However, this cannot explain the differences in their spectral energy 
distributions (SEDs); most of the RBLs, like the core-dominated quasars (CDQs), 
have a synchrotron peak in the NIR/optical (low-energy peaked 
BL Lacs, LBLs) while the majority of the XBLs have this peak in the 
UV/soft X-ray (high-energy peaked BL Lacs, HBLs). 
Higher electron Lorentz factors and/or magnetic fields in HBLs have been 
suggested to explain the differing peaks \citep{Sambruna96}. 
The CDQs, LBLs and HBLs form an SED sequence of increasing synchrotron peak 
frequency.

We explored if the {\bf B} field geometry follows an 
analogous trend using observations of HEAO-1 XBLs using the NRAO 
\footnote{The National Radio Astronomy Observatory is a facility of the
National Science Foundation operated under cooperative agreement by
Associated Universities, Inc.} Very Large Baseline Array at 5 GHz
\citep[Kharb et al., in prep]{Kollgaard96}.  Contrary to the 
tendency observed for the LBLs, the HBLs showed predominantly longitudinal 
{\bf B} fields relative to the local jet direction ({\it eg.,} Fig.~2a).
Fig.~2b shows a counter example: the HBL 1727+502, which has
a transverse ${\bf B}$ field ``spine" with longitudinal ${\bf B}$ field
at the jet edges.

There appears to be no straightforward interpretation of the {\bf B} 
field geometry trends in the beamed FRIs versus the beamed FRIIs 
within the simple US. Also, the ${\bf B}$ field geometries in CDQs, LBLs and 
HBLs do not reflect the SED synchrotron peak sequence.

\begin{figure}[h]
\centerline{
(a)
\includegraphics[width=6.6cm,height=6.6cm]{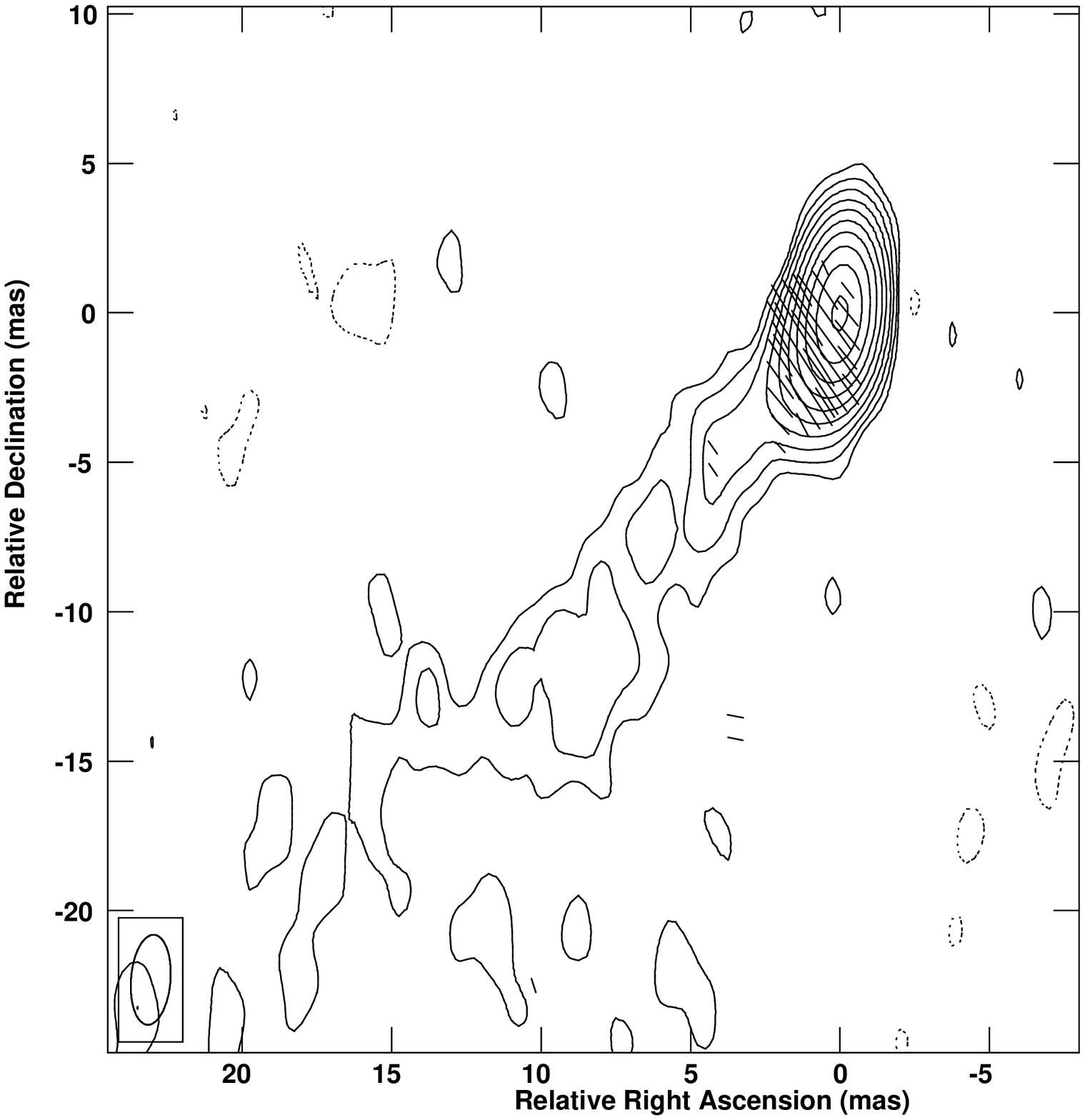}
(b)
\includegraphics[width=6.6cm,height=6.6cm]{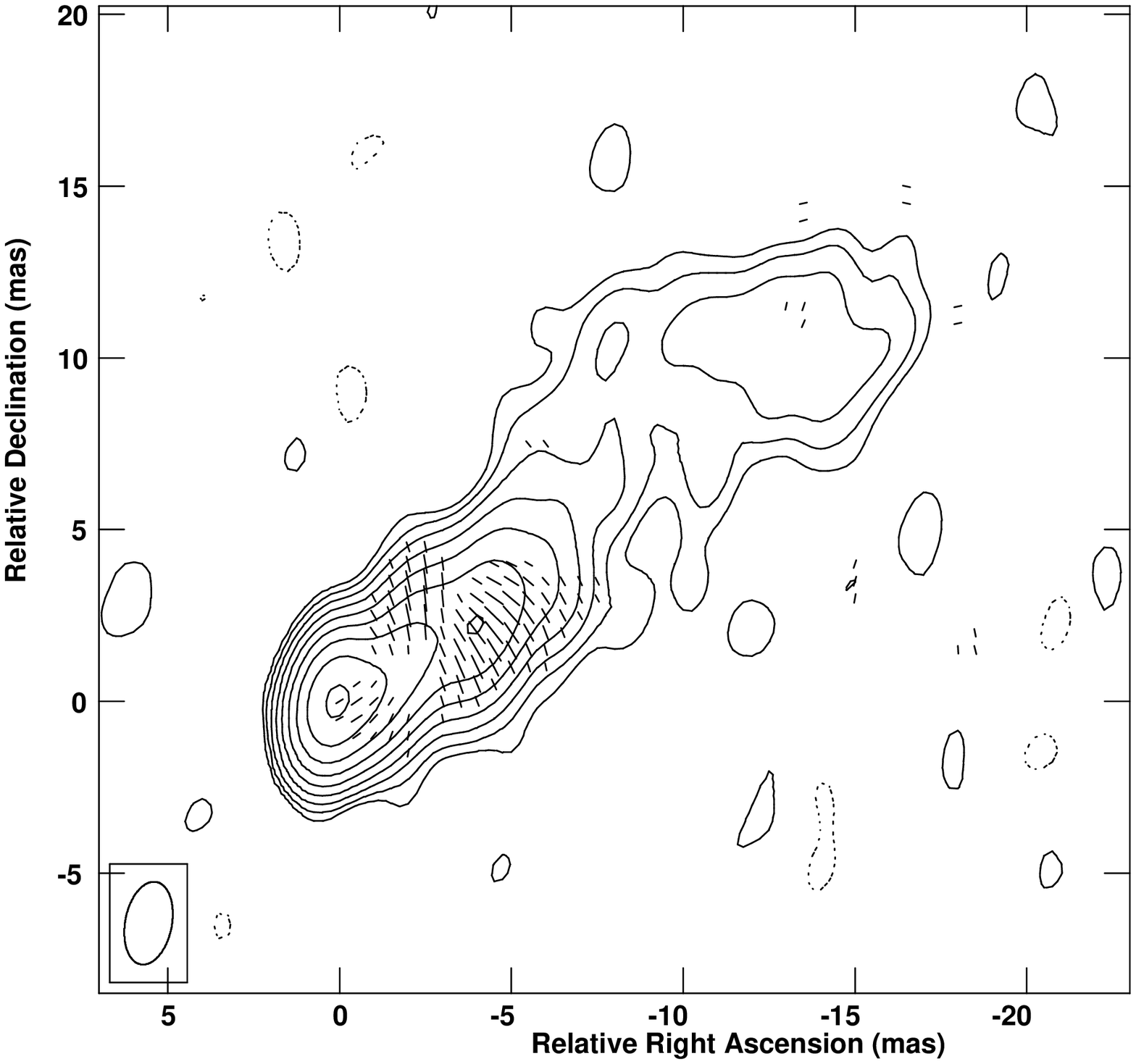}}
\caption{Total intensity VLBI maps of the HBLs (a) 1215+303 and (b) 
1727+502 at 5 GHz with polarization vectors superimposed. Contours are (a)
-0.17, 0.17, 0.35, 0.70, 1.40, 2.80, 5.60, 11.20, 22.50, 45 and 90 percent 
of the peak brightness of 231.1 mJy beam$^{-1}$, $\chi$ vectors: 1 mas =
1.3 mJy beam$^{-1}$ and (b)
-0.35, 0.35, 0.70, 1.40, 2.80, 5.60, 11.20,
22.50, 45 and 90 per cent of the peak brightness of 64.2 mJy beam$^{-1}$,
$\chi$ vectors: 1 mas = 1.8 mJy beam$^{-1}$.}
\end{figure}

\section{Conclusions}
\begin{enumerate}
\item All BLRGs that have been observed with the {\it HST} have detected 
pc-scale optical cores, consistent with the US.
\item The optical core luminosity correlates significantly with $R_{c}$ for 
the FRIs and BL~Lacs, as well as the FRI radio galaxies alone.
For the FRII population, only the
BLRGs and quasars show a significant correlation while 
the narrow-line objects do not, consistent with obscuration effects 
by a torus in the latter.
\item Relativistic beaming alone can account for the variation in the optical
cores of the FRI objects (beamed and unbeamed).
\item Modelling the optical emission as a combination of contributions from
a jet and a disk suggests that the accretion disks of FRII objects are more
luminous than those of FRI objects. This in turn suggests that the F--R 
division might be due to intrinsic differences on pc-scales.
\item The pc-scale jets of HBLs have predominantly longitudinal {\bf B} fields
w.r.t. the local jet direction, in contrast to the tendency for LBLs.
\item The ${\bf B}$ field geometries in CDQs, LBLs and HBLs do not reflect 
the SED synchrotron peak sequence of increasing frequency.
Thus, no simple conclusions can be drawn about the pc-scale {\bf B} field
geometries and their connection with the F--R divide. 
\end{enumerate}

\bibliographystyle{aa}
\bibliography{c_pkharb}
\end{document}